\documentclass{ifacconf}

\usepackage{graphicx}      
\usepackage{natbib}        

\usepackage{amsmath}
\usepackage{amssymb}
\usepackage{mathtools, cuted}
\usepackage{mathrsfs}
\usepackage{mleftright}
\usepackage{soul}
\usepackage{framed}
\usepackage{booktabs}
\usepackage{siunitx}
\usepackage{bbm}
\usepackage{tikz,pgfplots}
\usepackage{makecell}
\usepackage{url}

\usetikzlibrary{arrows,shapes,positioning}
\usetikzlibrary{tikzmark,calc}
\usetikzlibrary{arrows.meta}
\usetikzlibrary{decorations.pathreplacing}

\definecolor{myred}{rgb}{0.6350, 0.0780, 0.1840}
\definecolor{mygreen}{rgb}{0.4660, 0.6740, 0.1880}
\definecolor{myblue}{rgb}{0, 0.4470, 0.7410}

\newtheorem{definition}{Definition}

\DeclareMathOperator{\diag}{\mathrm{diag}}
\newcommand{\Norm}[1]{\mleft \Vert #1 \mright \Vert}
\newcommand{\lnorm}[1]{ \Vert #1  \Vert}

\mathtoolsset{showonlyrefs}

\usepackage[all=normal,wordspacing=tight,tracking=tight]{savetrees}

\begin{document}
\begin{frontmatter}

\title{Estimating Hormone Concentrations in the Pituitary-Thyroid Feedback Loop from Irregularly Sampled Measurements\thanksref{footnoteinfo}} 

\thanks[footnoteinfo]{This project has received funding from the European Research Council (ERC) under the European Union’s Horizon 2020 research and innovation programme (grant agreement No 948679). © 2026 the authors. This work has been accepted to IFAC for publication under a Creative Commons Licence CC-BY-NC-ND.}

\author[Hannover]{Seth Siriya} 
\author[Hannover]{Tobias M. Wolff}
\author[Hannover]{Isabelle Krauss}
\author[Hannover]{Victor G. Lopez}
\author[Hannover]{Matthias A. M{\"u}ller}

\address[Hannover]{Leibniz University Hannover, Institute of Automatic Control, Germany, (e-mail: {siriya, wolff, krauss, lopez, mueller}@irt.uni-hannover.de).}

\begin{abstract}
Model-based control techniques have recently been investigated for the recommendation of medication dosages to address thyroid diseases. These techniques often rely on knowledge of internal hormone concentrations that cannot be measured from blood samples. Moreover, the measurable concentrations are typically only obtainable at irregular sampling times. In this work, we empirically verify a notion of sample-based detectability that accounts for irregular sampling of the measurable concentrations on two pituitary-thyroid loop models representing patients with hypo- and hyperthyroidism, respectively, and include the internal concentrations as states. We then implement sample-based moving horizon estimation for the models, and test its performance on virtual patients across a range of sampling schemes. 
Our study shows robust stability of the estimator across all scenarios, and that more frequent sampling leads to less estimation error in the presence of model uncertainty and misreported dosages.

\end{abstract}

\begin{keyword}
Pituitary-thyroid feedback loop, hypothyroidism, hyperthyroidism, Hashimoto's thyroiditis, Graves' disease, irregular sampling, moving horizon estimation, detectability.
\end{keyword}

\end{frontmatter}

\section{Introduction}

\begin{figure} 
	\centering
	\begin{tikzpicture}[scale=1]
		\draw[line width = .5mm, color=mygreen](3,0.16) -- (7,0.16);
		\draw[line width = .5mm, color=mygreen,->](7,0.16) -- (7,-0.59);
		\draw[line width = .5mm, color=myblue](3,-0.16) -- (3.525,-0.16);
		\draw[line width = .5mm, color=myblue,->](3.5,-0.65) -- (4,-0.65);
		\node[align=center] at (3.25,.36) {$T_3$};
		\draw[line width = .5mm, rounded corners] (1,-0.33) rectangle (3,.33);
		\node[align=center] at (2,0) {Thyroid};
		\draw[line width = .5mm,->, color=mygreen](6,-0.65) -- (6.9,-0.65);
		\draw[line width = .5mm, rounded corners] (4,-0.98) rectangle (6,-0.33);
		\node[align=center] at (5,-0.65) {D1 \& D2};
		\node[align=center] at (6.5,-0.88) {$T_3$};
		\draw[line width = .5mm,->, color=mygreen](7.1,-0.65) -- (7.5,-0.65);
		\draw[line width = .5mm] (7,-0.65) circle (0.1);
		\draw[line width = .3mm](6.9,-0.65) -- (7.1,-0.65);
		\draw[line width = .3mm](7,-0.59) -- (7,-0.72);
		\node[align=center] at (3.2,-0.65) {$T_4$};
		\draw[line width = .5mm, color=myblue](3.5,-0.16) -- (3.5,-1.32);
		\draw[line width = .5mm,->, color=myblue](3.5,-1.3) -- (3,-1.3);
		\draw[line width = .5mm, rounded corners] (1,-1.63) rectangle (3,-0.98);
		\node[align=center] at (2,-1.3) {Pituitary};
		\node[align=center] at (1.4,-0.65) {$TSH$};
		\draw[line width = .5mm,->, color=myred](2,-0.98) -- (2,-0.33);
		\draw[line width = .5mm,->](0,-1.3) -- (1,-1.3);
		\node[align=center] at (0.2,-1.1) {$TRH$};
	\end{tikzpicture}
	\caption{Simple diagram of PT loop from \cite{wolff2025modeling}.}
	\label{fig:pt-loop}
\end{figure}
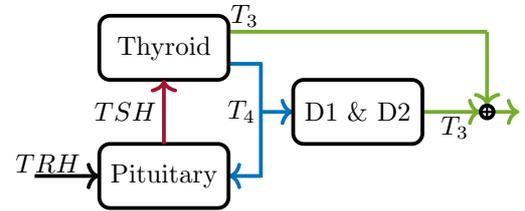

The pituitary-thyroid (PT) feedback loop---illustrated in Fig. \ref{fig:pt-loop}---is a key part of the endocrine system. Its functioning is as follows \citep{Gardner2004}. The pituitary gland secretes thyroid-stimulating hormone (TSH), stimulating production of the hormones triiodothyronine (T\textsubscript{3}) and thyroxine (T\textsubscript{4}) by the thyroid gland, the latter of which is further converted into T\textsubscript{3} by 5'-deiodinase type I (D1) and 5'-deiodinase type II (D2). The production of TSH is then inhibited by T\textsubscript{4} via the pituitary, closing the loop, while thyrotoropin-releasing hormone (TRH) produced by the hypothalamus stimulates TSH production.

Various diseases are associated with dysfunction of this loop. Hashimoto's thyroiditis is a disorder causing insufficient activity in the thyroid, leading to \textit{hypothyroidism}. It is typically treated via the oral intake of levothyroxine (L-T\textsubscript{4}) and sometimes levothyronine (L-T\textsubscript{3}), which are synthetic replacements for T\textsubscript{4} and T\textsubscript{3}, respectively. In contrast, Graves' disease is a condition causing antibodies to overstimulate thyroid activity, leading to \textit{hyperthyroidism}. It is often treated via the oral intake of antithyroid agents such as methimazole (MMI), inhibiting thyroid activity.

Historically, trial-and-error has been used to prescribe dosages based on measured T\textsubscript{4}, T\textsubscript{3}, and TSH concentrations from blood samples. This approach can take a long time for restoration of normal thyroid function, during which patients may be exposed to high dosages. Recently, systematic approaches have been investigated combining models with algorithmic tools, including control-theoretic methods.
In \cite{yang2021unified}, a PT loop model is proposed and analyzed.
A model considering oral intake of L-T\textsubscript{4} is used to design pole-placement and model predictive control (MPC) schemes for dosage recommendation in \cite{sharma2023automatic,sharma2024model}.
These works do not account for the influence of T\textsubscript{3}, which is known to be important for quality of life \citep{dietrich2012tsh, hoermann2017dual, hoermann2015homeostatic}.
On the other hand, \cite{wolff2022optimal} extend a PT loop model that accounts for T\textsubscript{3} from \citep{dietrich2002hypophysen,berberich2018mathematical} to consider oral intake of L-T\textsubscript{4} and L-T\textsubscript{3}, then apply nonlinear MPC for dosage selection. 
In the hyperthyroidism case, \cite{theiler2022mathematical} combine a proportional-integral controller with a PT loop model for MMI dosage recommendation, but neither T\textsubscript{3} nor oral intake of MMI are considered. In contrast, the nonlinear model in \cite{wolff2025modeling} considers these, and is combined with MPC for dosage recommendation. However, the models from \cite{wolff2022optimal,wolff2025modeling} involve internal concentrations of T\textsubscript{4} in the thyroid alongside T\textsubscript{3} and TSH in the pituitary as states, which \textit{cannot} be readily measured from blood samples, motivating the use of state estimators. Although peripheral T\textsubscript{4}, T\textsubscript{3}, and TSH concentrations \textit{can} be measured from blood samples, testing can only occur irregularly.


To address the state estimation problem for nonlinear systems, the extended and unscented Kalman filter \citep{stengel1994optimal,julier2004unscented} are popular observers due to their computational efficiency, but often produce inaccurate estimates for highly nonlinear models and unmodeled disturbances \citep{rawlings2020model}. Moving horizon estimation (MHE) overcomes these issues. It formulates the estimation task as an optimization problem over a finite horizon of past output measurements and inputs, given a known system model. Under a detectability notion known as incremental input/output-to-state stability (i-IOSS), robust stability of optimization-based observers such as MHE can be guaranteed \citep{allan2021robust, ji2015robust, knufer2023nonlinear, schiller2023lyapunov, schiller2024robust}.
However, MHE requires constant, regular sampling of output measurements, which is not reasonable in the PT loop since blood samples are collected sparsely and irregularly. Recently, sample-based versions of i-IOSS and MHE that handle irregular measurements have been developed in \cite{krauss2025detectability} and \cite{krauss2025sample}, respectively.

Our contributions are as follows. Firstly, we obtain approximate discrete-time (DT) models of patients with hypo- and hyperthyroidism that are suitable for use by optimization-based observers by adapting the models from \cite{wolff2022optimal,wolff2025modeling}, and empirically verify sample-based i-IOSS with respect to some outpatient and inpatient-relevant sampling schemes.
Secondly, we implement sample-based MHE based on \cite{krauss2025sample} on these DT systems. We test this on two virtual patients with hypo- and hyperthyroidism, and observe that our method is robustly stable across the sampling schemes, with performance improvements when sampling more frequently.

\paragraph*{Notation}
Given $a,b \in \mathbb{Z}$, define $\mathbb{Z}_a := \mathbb{Z} \cap [a,\infty)$, and $\mathbb{Z}_a^b:=\mathbb{Z}\cap[a,b]$.
Given a set $\mathcal{X}$, $\mathcal{X}^{\infty}$ denotes the set of all sequences $\{x_i\}_{i=0}^{\infty}$ satisfying $x_i \in \mathcal{X}$ for all $i \geq 0$. 
Given a vector $x \in \mathbb{R}^n$ and a matrix $P \succ 0$, $\Norm{x}_P = \sqrt{x^{\top} P x}$ denotes the weighted 2-norm.
Given $t_2 > t_1 \geq 0$, a vector space $\mathcal{S}$, and a signal $s : [0, \infty) \rightarrow \mathcal{S}$, $s_{[t_1,t_2]}(\tau):= s(t_1 + \tau)$ for $\tau \in [0,t_2 - t_1]$ is the truncation of $s$ over $[t_1,t_2]$.
The indicator function of a set $A$ is denoted by $\mathbf{1}_A$. 

\section{System Modeling}
\label{sec:system}

Continuous-time (CT) models of patients treated for hypo- and hyperthyroidism are provided in Sec.~\ref{sec:model-ct}.
In Sec.~\ref{sec:adt-models}, we introduce the corresponding DT models used for MHE.

\subsection{Continuous-time models} \label{sec:model-ct}
The CT model for hypothyroidism treatment is
\begin{align}
	\dot{x} = f_{\text{hypo}}(x,u,w). \label{eqn:f-hypo}
\end{align}
The system state is $x = \begin{bmatrix} T_{4,th} & T_4 & T_{3p} & T_{3c} & TSH & TSH_c \end{bmatrix}^{\top} \in \mathbb{R}_{\geq 0}^6$, where $T_4$ ($10^{-7}$ mol/l), $T_{3p}$ ($10^{-9}$ mol/l), and $TSH$ (mIU/l) are the peripheral concentrations of T\textsubscript{4}, T\textsubscript{3}, and TSH, respectively, $T_{4,th}$ ($10^{-12}$ mol/l) is the internal concentration of T\textsubscript{4} in the thyroid, and $T_{3c}$ ($10^{-8}$ mol/l), $TSH_c$ (mIU/l), are the internal concentrations of T\textsubscript{3}, TSH, in the pituitary, respectively.
Moreover, $u(t) = [u_{L\text{-}T_3}(t) \ u_{L\text{-}T_4}(t) ]^{\top} \in \mathbb{R}_{\geq 0}^2$ is a known, time-varying input, where $u_{L\text{-}T_3}(t)$ and $u_{L\text{-}T_4}(t)$ describe the time-dependent absorption of the medications L-T\textsubscript{3} and L-T\textsubscript{4} (mol/l/s), and are computable given historical knowledge of L-T\textsubscript{3} and L-T\textsubscript{4} dosages taken by a patient \eqref{eqn:lt3}-\eqref{eqn:lt4} in Appendix~\ref{sec:append-hypo}.
The model $f_{\text{hypo}}$ corresponds to the right-hand sides of \eqref{eqn:hypo-dynamics-1}--\eqref{eqn:hypo-dynamics-6}. Table~\ref{tab:params-fixed} in Appendix~\ref{sec:params} contains the model parameters, except $G_{T,co}$---the scaling factor for the thyroid hormone production rate---is set to $0.1$, representing underactivity.
It is obtained by modifying the model in \cite{wolff2022optimal} to consider the process noise $w = \begin{bmatrix} w_{G_{D1}} & w_{G_{T3}} & w_{TRH} & w_{L\text{-}T_3} & w_{L\text{-}T_4}\end{bmatrix}^{\top} \in \mathbb{R}^5$. We focus on describing our modifications.
In particular, $w_{G_{D1}}$, $w_{G_{T3}}$, $w_{TRH}$, $w_{L\text{-}T_3}$, and $w_{L\text{-}T_4}$ capture uncertainty in the crucial parameters $G_{D1}$, $G_{T3}$, and $TRH$, alongside $u_{L\text{-}T_3}$ and $u_{L\text{-}T_4}$ due to misreported dosages. Compared to \cite{wolff2022optimal}, $f_{\text{hypo}}$ is obtained by multiplying $G_{D1}$, $G_{T3}$, $TRH$, 
$u_{L\text{-}T_4}$, and $u_{L\text{-}T_3}$, with $(1 + w_{G_{D1}})$, $(1 + w_{G_{T3}})$, $(1 + w_{TRH})$, $(1 - w_{L\text{-}T_4})$, and $(1 - w_{L\text{-}T_3})$, respectively.

The output model for the measured peripheral T\textsubscript{4}, T\textsubscript{3} and TSH from blood samples is $y = h(x,v) := [(T_4 + v_{T_4}) \quad (T_{3p} + v_{T_{3p}}) \quad (TSH + v_{TSH}) ]^{\top} \in \mathbb{R}^3$, where $v = \begin{bmatrix} v_{T_4} & v_{T_{3p}} & v_{TSH} \end{bmatrix}^{\top} \in \mathbb{R}^3$ is measurement noise.

The CT model for hyperthyroidism is
\begin{align}
	\dot{x} = f_{\text{hyper}}(x,u,w). \label{eqn:f-hyper}
\end{align}
The state is $x \in \mathbb{R}_{\geq 0}^7$, which compared to hypothyroidism includes $MMI_{th}$, the concentration of MMI in the thyroid ($10^{-5}$ mol/l). The known input is $u(t) = u_{MMI}(t) \in \mathbb{R}_{\geq 0}$, describing the absorption of $MMI$ from the plasma to the thyroid (mol/l/s), and the process noise is $w = \begin{bmatrix} w_{G_{D1}} & w_{G_{T3}} & w_{TRH} & w_{MMI}\end{bmatrix}^{\top} \in \mathbb{R}^4$, where $w_{MMI}$ captures uncertainty in $u_{MMI}$ from misreported dosages. Development of $f_{\text{hyper}}$ is described in more depth within Appendix~\ref{sec:append-hyper}.

\subsection{Approximate Discrete-Time Models}
\label{sec:adt-models}
In this section, we describe the DT models based on Sec.~\ref{sec:model-ct} that will be used for sample-based MHE.

We start with the hypothyroidism case.
Denote the solution to \eqref{eqn:f-hyper} at time $t$, initialized from a state $x \in \mathcal{X}_{\text{hypo}} \subseteq \mathbb{R}^6$ at time $0$, and driven by a known signal $u:[0,t]\rightarrow \mathbb{R}_{\geq 0}$ and fixed process noise $w \in \mathcal{W}_{\text{hypo}} \subseteq \mathbb{R}^5$, by $\phi_{\text{hypo}}(t,x,u(\cdot),w)$.
Let $F_{\text{hypo}}$ be a discretized model of \eqref{eqn:f-hypo} satisfying
$ F_{\text{hypo}}(x,u(\cdot),w) \approx \phi_{\text{hypo}}(T_d, x, u(\cdot), w) $
for $x \in \mathcal{X}_{\text{hypo}}$, $w \in \mathcal{W}_{\text{hypo}}$, and $u(\cdot) \in \{ U_{[kT_d,(k+1)T_d]} \mid U(\cdot) \in \mathbb{U}_{L\text{-}T}, k \in \mathbb{Z}_{0} \}$.\footnote{We model $u(\cdot)$ as a CT signal since it can be symbolically expressed as a function of time via \eqref{eqn:lt3}--\eqref{eqn:lt4}, which the CVODES integrator supports. However, $w$ is fixed since it will be used as a decision variable when implementing MHE in Sec.~\ref{sec:sb-mhe}.} Here, $T_d$ is the discretization period, which is chosen as $T_d = 8$ \si{\hour} in this work.
Moreover, $\mathbb{U}_{L\text{-}T}$, is the set of possible $u_{L\text{-}T_3}$ and $u_{L\text{-}T_4}$ with up to $30$ \si{\micro\gram} of L-T\textsubscript{3} and $400$ \si{\micro\gram} of L-T\textsubscript{4} taken daily, as defined in \eqref{eqn:U-hypo} in Appendix~\ref{sec:append-hypo}.
This can be obtained via various numerical integration methods (e.g., Euler and Runge-Kutta), but we utilize the CVODES integrator in the SUNDIALS suite from \cite{hindmarsh2005sundials} due to the stiffness of \eqref{eqn:f-hypo}.
We can then formulate the following DT system:
\begin{align}
	x_{k+1} = F_{\text{hypo}}(x_k,u_k(\cdot),w_k), \quad y_k = h(x_k,v_k),
	\label{eqn:hypo-dt-system}
\end{align}
with $x_k \in \mathcal{X}_{\text{hypo}}$, $y_k \in \mathcal{Y}_{\text{hypo}} \subseteq \mathbb{R}^3$, $w_k \in \mathcal{W}_{\text{hypo}}$, $v_k \in \mathcal{V} \subseteq \mathbb{R}^{3}$, and $u_k(\cdot) = U_{[kT_d,(k+1)T_d]}$ for $k \in \mathbb{Z}_0$, where $U(\cdot) \in \mathbb{U}_{L\text{-}T}$. 
We consider $\mathcal{X}_{\text{hypo}} = [0.2,0.6] \times [0.1,1.4] \times [0.8,3.1] \times [0.1,1.3] \times [1.4,6] \times [1.5,6.3]$ ($10^{-12}$ mol/l, $10^{-7}$ mol/l, $10^{-9}$ mol/l, $10^{-8}$ mol/l, mIU/l, mIU/l), corresponding to low-to-normal thyroid activity, $\mathcal{W}_{\text{hypo}} = [-0.1,0.1]^2 \times [-0.3,0.3] \times \{0\} \times [0, 1] $,\footnote{We consider $w_{L\text{-}T_3} \in \{0\}$ since we will only simulate a virtual patient medicated with $L$-$T_4$, which is consistent with treatment guidelines in \cite{jonklaas2014guidelines}. However, the general method is still applicable with $w_{L\text{-}T_3} \in [0,1]$ for misreported $L$-$T_3$ dosages.} $\mathcal{V}= [-0.1 h(x_{\text{ss}},0),$ $0.1 h(x_{\text{ss}},0)]$ ($10^{-7}$ mol/l, $10^{-9}$ mol/l, mIU/l), and $\mathcal{Y}_{\text{hypo}}=\mathcal{X}_{\text{hypo}}\oplus \mathcal{V}$, where $x_{\text{ss}}=\begin{bmatrix} 3.10 & 1.17 & 2.71 & 1.12 & 1.87 & 1.99 \end{bmatrix}^{\top}$ ($10^{-12}$ mol/l, $10^{-7}$ mol/l, $10^{-9}$ mol/l, $10^{-8}$ mol/l, mIU/l, mIU/l) is the steady-state solution of \eqref{eqn:f-hypo} when $G_{T,co}=1$ for a healthy individual. We also denote $\mathbb{W}_{\text{hypo}} = \mathcal{W}_{\text{hypo}} \times \mathcal{V}$.

The hyperthyroidism case follows similarly, so we just describe the key differences. 
Denote the discretized dynamics by $F_{\text{hyper}}$.
We formulate the following DT system:
\begin{align}
	x_{k+1} = F_{\text{hyper}}(x_k,u_k(\cdot),w_k), \ y_k = h(x_k,v_k),
	\label{eqn:hyper-dt-system}
\end{align}
with $x_k \in \mathcal{X}_{\text{hyper}} \subseteq \mathbb{R}^7$, $y_k \in \mathcal{Y}_{\text{hyper}} \subseteq \mathbb{R}^3$, $w_k \in \mathcal{W}_{\text{hyper}} \subseteq \mathbb{R}^4$, $v_k \in \mathcal{V} \subseteq \mathbb{R}^3$, and $u_k(\cdot) = U_{[kT_d,(k+1)T_d]}$ for $k \in \mathbb{Z}_{0}$, where $U(\cdot) \in \mathbb{U}_{MMI}$ and $\mathbb{U}_{MMI}$ is the set of possible $u_{MMI}(\cdot)$ with up to $35$ \si{\milli\gram} of MMI taken daily (defined in \eqref{eqn:U-mmi} in Appendix~\ref{sec:append-hyper}).
We consider $\mathcal{X}_{\text{hyper}} = [0.2,17] \times [0.4,5] \times [1.2,11] \times [0.4,4.5] \times [0.6,3.5] \times [0.7,3.5] \times [0,5]$ ($10^{-12}$ mol/l, $10^{-7}$ mol/l, $10^{-9}$ mol/l, $10^{-8}$ mol/l, mIU/l, mIU/l, $10^{-5}$ mol/l), corresponding to normal--high thyroid activity, $\mathcal{W}_{\text{hyper}} = [-0.1,0.1]^2 \times [-0.3,0.3] \times [0, 1]$, and $\mathcal{V}$ as before. Moreover, $\mathcal{Y}_{\text{hyper}} = 
\mathcal{X}_{\text{hyper}} \oplus \mathcal{V}$, and $\mathbb{W}_{\text{hyper}} = \mathcal{W}_{\text{hyper}} \times \mathcal{V}$.

For brevity, throughout this work, $(F,\mathcal{X},\mathcal{Y},\mathbb{U},\mathbb{W})$ is a placeholder for $(F_{\text{hypo}},\mathcal{X}_{\text{hypo}},\mathcal{Y}_{\text{hypo}},\mathbb{U}_{L\text{-}T},\mathbb{W}_{\text{hypo}})$ in the hypothyroidism case, and $(F_{\text{hyper}},$ $\mathcal{X}_{\text{hyper}},\mathcal{Y}_{\text{hyper}},\mathbb{U}_{MMI},\mathbb{W}_{\text{hyper}})$ for hyperthyroidism.

\section{Verifying Detectability}
\label{sec:detectability}

We first recall the definition of a sampling set and sample-based exponential i-IOSS from \cite{krauss2025sample}.

\begin{definition}[Sampling set]
	Consider a sampling interval sequence $\{\delta_i\}_{i =1}^{\infty}$ satisfying $\delta_i \in \mathbb{Z}_{0}$ for $i \in \mathbb{N}$ and $\sup_i \delta_i < \infty$. Then, $K_i := \{\sum_{k=i}^{j+i-1} \delta_k \mid j \in \mathbb{N} \}$ is the associated infinite set of time instances from index $i$ on, and $K := \{ K_i \mid i \in \mathbb{N}\}$ is the corresponding sampling set.
\end{definition}

\begin{definition}[Sample-based exponential i-IOSS]
	\label{def:sb-iioss}
	 \ \\Consider a sampling set $K$.
	Then, the system $(F,h)$ is sample-based exponentially i-IOSS with respect to 
	$K$
	if there exist $\bar{P}_1,\bar{P}_2 \succ 0$, $\bar{Q},\bar{R} \succeq 0$ and $\bar{\eta} \in [0,1)$ such that for all  $x_0,\tilde{x}_0 \in \mathcal{X}$, $\{\omega_i\}_{i=0}^{\infty},\{\tilde{\omega}_i\}_{i=0}^{\infty} \in \mathbb{W}^{\infty}$, and $U(\cdot) \in {\mathbb{U}}$,
	\begin{align}
		\Norm{x_k - \tilde{x}_k}_{\bar{P}_1}^2 &\leq  \Norm{x_0 - \tilde{x}_0}_{\bar{P}_2}^2 \bar{\eta}^k + \sum_{j = 0}^{k-1} \bar{\eta}^{k - j - 1} \Norm{\omega_{j} - \tilde{\omega}_{j}}_{\bar{Q}}^2 \\
		&+ \sum_{j \in \mathbb{Z}_{0}^{k-1} \cap K_i} \bar{\eta}^{k - j - 1}\lnorm{y_{j} - \tilde{y}_{j}}_{\bar{R}}^2 \label{eqn:sb-iioss-ineq}
	\end{align}
	holds for all $k \geq 0$ and $K_i \in K$, where $\omega_k = [w_k^{\top} \ v_k^{\top}]^{\top}$, $\tilde{\omega}_k = [\tilde{w}_k^{\top} \ \tilde{v}_k^{\top}]^{\top}$, $u_k(\cdot) = U_{[kT_d,(k+1)T_d]} (\cdot)$,  $\{(x_k,y_k)\}_{k=0}^{\infty}$ is the state-output trajectory associated with the initial state-input-noise triple $(x_0,\{u_k(\cdot)\}_{k=0}^{\infty},\{\omega_k\}_{k=0}^{\infty})$, and $\{(\tilde{x}_k,\tilde{y}_k)\}_{k=0}^{\infty}$ is associated with $(\tilde{x}_0,\{u(\cdot)_k\}_{k=0}^{\infty},\{\tilde{\omega}_k\}_{k=0}^{\infty})$.
\end{definition}

Regular i-IOSS has become the standard detectability notion for nonlinear MHE in recent years \citep{allan2021robust,schiller2023lyapunov}, but does not support irregular sampling. Instead, \cite{krauss2025sample} showed that under the assumption of \textit{sample-based} i-IOSS in Def.~\ref{def:sb-iioss}, robust global exponential stability can be established for the estimation error resulting from a sample-based version of MHE considering irregular sampling (described in Sec.~\ref{sec:sb-mhe}). Thus, verification of sampled-based i-IOSS is of interest, in order to evaluate suitability of the DT models from Sec.~\ref{sec:adt-models} for observer design with irregular measurements.

\begin{table}[htbp]
	\caption{Sequences of sampling intervals.}
	\label{tab:samp-sequence-outpatient}
	{\centering
		\begin{tabular}{lccc}
			\toprule
			Notation & Defining rule & \makecell{Interval} \\
			\midrule
			$\{\delta_{i}^a\}_{i \in \mathbb{N} }$ & $\delta^a_i = i$ & 8 hours \\
			$\{\delta_{i}^b\}_{i \in \mathbb{N} }$ & $\delta_i^b = \mathbf{1}_{[2,\infty)}(i)3 \delta^0_i$ & 1-1.5 weeks \\
			$\{\delta_{i}^c\}_{i \in \mathbb{N} }$ & $\delta_{i}^c = \mathbf{1}_{[2,\infty)}(i) (\delta^b_{2(i-1)} + \delta^b_{2(i-1) + 1})$  & 2-3 weeks\\
			$\{\delta_{i}^d\}_{i \in \mathbb{N} }$ & $\delta_{i}^d = \mathbf{1}_{[2,\infty)}(i)(\delta^c_{2(i-1)} + \delta^c_{2(i-1) + 1})$ & 4-6 weeks \\		
			\bottomrule
		\end{tabular} \par}
	$\{\delta^0_j\}_{j=1}^{\infty}$ is defined as $\delta^0_j := 10 - | (j-2) \mod (6 - 3) |$ for $j \in \mathbb{N}$, producing the periodic sequence $\{8, 7, 8, 9, 10, 9, 8, 7, 8, \hdots\}$.
\end{table}

\begin{figure}
	\centering
	\includegraphics[width=0.5\textwidth]{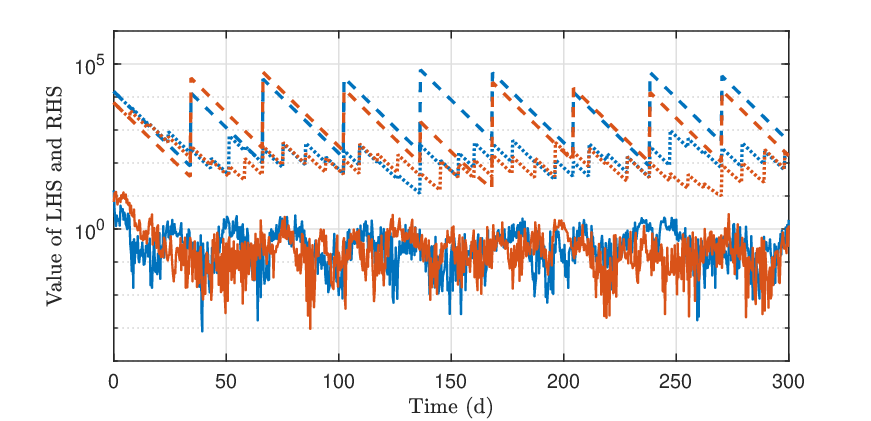}
	\caption{LHS and RHS of \eqref{eqn:sb-iioss-ineq} for two sampled trajectory pairs in the hypothyroidism scenario. The colors red and blue correspond to each pair. For each pair, the solid line corresponds to the LHS, and the dotted and dashed lines correspond to the RHS with sampled time instances in $K_2^b$ and $K_2^d$, respectively.}
	\label{fig:sb-iioss-hypo}
\end{figure}

To empirically verify sample-based i-IOSS for the hypothyroidism model, we consider the sampling interval sequences from Table~\ref{tab:samp-sequence-outpatient}. 
Sequences $\{\delta_{i}^b\}_{i \in \mathbb{N}}$, $\{\delta_{i}^c\}_{i \in \mathbb{N}}$, $\{\delta_{i}^d\}_{i \in \mathbb{N}}$, correspond to outpatient settings with sparse, irregular sampling, and $\{\delta_{i}^a\}_{i \in \mathbb{N}}$ corresponds to an inpatient setting with frequent, regular sampling.
We simulate 19800 trajectory pairs over $300$ days, with initial states and noise sampled uniformly over $\mathcal{X}_{\text{hypo}}$ and $\mathbb{W}_{\text{hypo}}$, and known inputs from $\mathbb{U}_{L\text{-}T}$ obtained by uniformly sampling the L-T\textsubscript{4} dosage each day over $[0,40]$ \si{\micro\gram}. 
We denote $K^l_i$ as the infinite set of time instances from interval $i \in \mathbb{N}$, and $K^l$ as the sampling set, both generated by  $\{\delta_i^l\}_{i=1}^{\infty}$ for $l \in \{a,b,c,d\}$.
In the case of $K^a$, we empirically verify that \eqref{eqn:sb-iioss-ineq} holds across all 19800 pairs over $300$ days by choosing $\bar{P}_1 = I_6$, $\bar{P}_2 = 2\cdot10^3 I_6$, $\bar{\eta}=0.95$, $\bar{Q} = 0.5 I_{8}$, and $\bar{R} = 0.5 I_3$. In the case of $K^b$, $K^c$, and $K^d$, \eqref{eqn:sb-iioss-ineq} is satisfied setting $\bar{R}$ as $5\cdot 10^2 I_3$, $5\cdot 10^3 I_3$, and $10^5 I_3$, respectively. We plot the LHS and RHS of \eqref{eqn:sb-iioss-ineq} for two random trajectory pairs in Fig.~\ref{fig:sb-iioss-hypo}, for the set of time instances $K_2^b$ and $K_2^d$.\footnote{We illustrate $K^l_2$, $l \in \{b,d\}$, rather than the apparent choice $K^l_1$ since $K^l_2 \subseteq K^l_1$, and therefore if \eqref{eqn:sb-iioss-ineq} holds for $K^l_2$, it holds for $K^l_1$. However, the trend in Fig.~\ref{fig:sb-iioss-hypo} is consistent across all $K^l_i$, $i \in \mathbb{N}$.}
The RHS typically decreases, but jumps at sampled time instances, such that the RHS remains higher than the LHS. 

For hyperthyroidism, we empirically verify that \eqref{eqn:sb-iioss-ineq} holds when choosing  $\bar{P}_1 = I_7$, $\bar{P}_2 = 2 \cdot 10^3 I_7$, $\bar{\eta}=0.96$, $\bar{Q} = 5 I_{7}$, and $\bar{R} = 20 I_3$ for $K_2^a$, and with $\bar{R}$ increased to $3\cdot10^3 I_3$, $5 \cdot 10^4 I_3$, and $2 \cdot 10^6 I_3$, for $K_2^b$, $K_2^c$, and $K_2^d$, respectively.

\section{Sample-Based MHE}
\label{sec:sb-mhe}

In this section, we describe sample-based MHE for estimating hormone concentrations with irregular sampling. It follows similarly to \cite{krauss2025sample}, except we consider a \textit{filtering} rather than a \textit{prediction} form, so the state estimate at a given time considers the corresponding measurement \citep[Ch.~4]{rawlings2020model}.\footnote{In practice, sampling delays can occur and vary based on a range of factors, including whether inpatient or outpatient care is being exercised. We consider instantaneously known measurements for simplicity, and leave considering sampling delays to future work.}

Let $K_s \subseteq \mathbb{Z}_{0}$ be the set of time instances where a measurement is available to the estimator, and define the horizon length as $M_k := \min \{k, M\}$.
At every time step $k \in K_s$ where a measurement is available, the following nonlinear program (NLP) is solved:
\begin{align}
	\min_{\hat{x}_{k-M_k \mid k},\hat{\omega}_{\cdot \mid k}} &J(\hat{x}_{k-M_k \mid k}, \hat{\omega}_{\cdot \mid k}, \hat{y}_{\cdot \mid k}, k) \\
	\text{s.t.} \ \hat{x}_{j+1 \mid k} &= F(\hat{x}_{j \mid k},u_j(\cdot),\hat{w}_{j\mid k}), \ j \in \mathbb{Z}_{k-M_k}^{k-1}, \\
	\hat{y}_{j \mid k} &= h(\hat{x}_{j \mid k}, \hat{v}_{j \mid k}), \ j \in \mathbb{Z}_{k-M_k}^{k}, \\
	\hat{w}_{j \mid k} &\in \mathcal{W}, \ \hat{y}_{j \mid k} \in \mathcal{Y}, \ j \in \mathbb{Z}_{k-M_k}^{k}, \\
	\hat{x}_{j\mid k} &\in \mathcal{X}, \ j \in \mathbb{Z}_{k-M_k}^{k}, \label{eqn:nlp}
\end{align}
where $u_j(\cdot) \in \mathbb{U}$. Here, $\hat{x}_{j \mid k}$ denotes the estimated state for time $j$ computed at time $k$, and $\hat{\omega}_{j\mid k} = [ \hat{w}_{j\mid k}^{\top} \ \hat{v}_{j\mid k}^{\top} ]^{\top}$ and $\hat{y}_{j \mid k} \in \mathcal{Y}$ denote the estimated noise and outputs analogously. 
The optimal state sequence that solves the NLP at time $k$ is denoted by $\hat{x}^{*}_{\cdot \mid k}$, and the optimal estimate at time $k$ is $\hat{x}_k := \hat{x}^*_{k \mid k}$. We consider the cost function
\begin{align}
	&J(\hat{x}_{k-M_k \mid k}, \hat{\omega}_{\cdot \mid k}, \hat{y}_{\cdot \mid k}, k) = 2\eta^{M_k} \lVert \hat{x}_{k - M_k \mid k} - {\tilde{x}}_{k - M_k}  \rVert_{P}^2 + \\
	& \sum_{j =k - M_k}^k \eta^{k - j} 2 \lVert \hat{\omega}_{j \mid k} \rVert_Q^2 + \sum_{j \in \mathbb{Z}_{k-M_k}^{k}\cap K_s} \eta^{k - {j}} \lVert \hat{y}_{j \mid k} - y_j \rVert_R^2. \label{eqn:cost-old}
\end{align}
Here, $\tilde{x}_k$ is the prior for the state used when solving the NLP, and is defined as $\tilde{x}_k := \hat{x}_k$ if $k > 0$, and $\tilde{x}_k:= \chi$ if $k = 0$, where $\chi \in \mathcal{X}$ is the chosen prior at time $k=0$.
Moreover, $P \succ 0$ is a matrix penalizing the difference between the estimated state at the beginning of the horizon and the prior, and $Q,R \succeq 0$ are matrices penalizing the magnitude of the estimated disturbance sequence, and the difference between estimated and measured outputs, respectively.
Finally, $\eta \in [0, 1)$ is a discount factor that determines the influence of past disturbances and measurements on the cost function.
In contrast to the above, when $k \not \in K_s$, such that there is no new information available, the estimate is instead computed via open-loop prediction by setting $\hat{x}_k := F(\hat{x}_{k-1},u_{k-1}(\cdot),0)$ if $k>0$, and $\hat{x}_0 := \chi$ if $k = 0$.

The sample-based MHE scheme is applied for both hypo- and hyperthyroidism, with the set of sampled time steps $K_s$ chosen to be $K^a_1$, $K^b_1$, $K^c_1$, and $K^d_1$. Note that $K_1^d \subseteq K^c_1 \subseteq K^b_1 \subseteq K^a_1$. 
In theory, the parameters $P,Q,R$, and $\eta$, can be chosen based on the matrices found in Sec.~\ref{sec:detectability} to guarantee robust stability of the estimator \citep{krauss2025sample}. However, we manually tune these parameters in this work, which is typically done to achieve good performance in practice. In particular, for hypothyroidism, we set $P = \diag( 1 , 0.1 , 1 , 1 , 1 , 1)$, $Q = \diag( 10 , 1 , 1 , 0 , 1 , 1000 , 1000,$ $100)$, and $R = \diag(500, 500, 100)$, with $\eta = 0.7$, and $M = 20$. 
For hyperthyroidism, we set $P = \diag(100 , 0.1 , 1 , 1 , 1 , 1 , 100 )$, $Q = \diag( 10 , 1 , 1 , 10 , 1000 ,$ $1000 , 100 )$, $R = \diag(  250 , 250 , 1000 )$, $\eta = 0.8$, and $M = 20$.

%
%

\section{Simulation Results}

We now implement the sample-based MHE scheme from Sec.~\ref{sec:sb-mhe} using CasADi \citep{andersson2019casadi} on simulated patients affected by hypo- and hyperthyroidism in Sec.~\ref{sec:sim-hypo} and \ref{sec:sim-hyper}, respectively, with the sets of sampled time instances $K_1^a$, $K_1^b$, $K_1^c$, and $K_1^d$.
\footnote{All code is available at \url{https://doi.org/10.25835/lyl7bocm}.}

\subsection{Hypothyroidism}
\label{sec:sim-hypo}

We consider a simulation scenario for hypothyroidism where the patient initially does not take any medication. Then, they follow a simple oral L-T\textsubscript{4} medication strategy based on the guidelines provided in \cite{jonklaas2014guidelines} starting from day 34. In particular, if the concentration of measured TSH is above the target range of $0.5$--$4$ mIU/l on day $34$, the patient initially takes a dosage of $136$ \si{\micro\gram}/day (corresponding to an 80 \si{\kilogram} human). Dose adjustments of $18.75$ \si{\micro\gram}/day are then made up/down every $4$--$6$ weeks, depending on whether measured TSH is above or below the target range, until it is reached. 
The corresponding signal $U(\cdot) \in \mathbb{U}_{L\text{-}T}$ is used when solving the MHE problem. However, we suppose the virtual patient forgets to take medication on days $39$--$43$ and $81$--$85$, which is not accounted for in the MHE scheme (i.e., $U(\cdot)$ does not reflect this), resulting in misreported dosages.

We simulated this scenario using the model from \eqref{eqn:f-hypo} with $U(\cdot)$ as the known input. However, we set $w_{L\text{-}T_4}(t) = 1 - U_{\text{true}}(t)/U(t)$ if $U(t) > 0$ and $w_{L\text{-}T_4}(t) = 0$ if $U(t)=0$ to represent misreported dosages, where $U_{\text{true}}(\cdot) \in \mathbb{U}_{L\text{-}T}$ is the hypothetical input if forgotten days are explicitly accounted for.\footnote{This choice of $w_{L\text{-}T_4}$ implies $U(t)(1-w_{L\text{-}T_4}(t)) = U_{\text{true}}(t)$.}
We set $w_{L\text{-}T_3}(t) = 0$ since the patient does not take $L$-$T_3$,  $w_{G_{D1}}(t)=w_{G_{T3}}(t)=0.1$ to represent elevated $G_{D1}$ and $G_{T3}$, and $w_{TRH}(t) = 0.3 \cos (\pi(\frac{t}{43200}-\frac{5}{12}))$ to capture the circadian rhythm of TRH.
Measurements are simulated by corrupting $T_4$, $T_{3p}$, and $TSH$ by a noise sampled uniformly over $\mathcal{V}$. The initial state is $x(0) = \begin{bmatrix} 0.49 & 0.18 & 0.92 & 0.18 & 5.14 & 5.48\end{bmatrix}^{\top}$, corresponding to the virtual patient's steady-state concentrations, and the prior is $\chi = \begin{bmatrix} 0.2 & 1.5 & 3 & 1.5 & 2 & 2 \end{bmatrix}^{\top}$, with the units ($10^{-12}$ mol/l, $10^{-7}$ mol/l, $10^{-9}$ mol/l, $10^{-8}$ mol/l, mIU/l, mIU/l).

\begin{figure}
	\centering
	\includegraphics[width=0.5\textwidth]{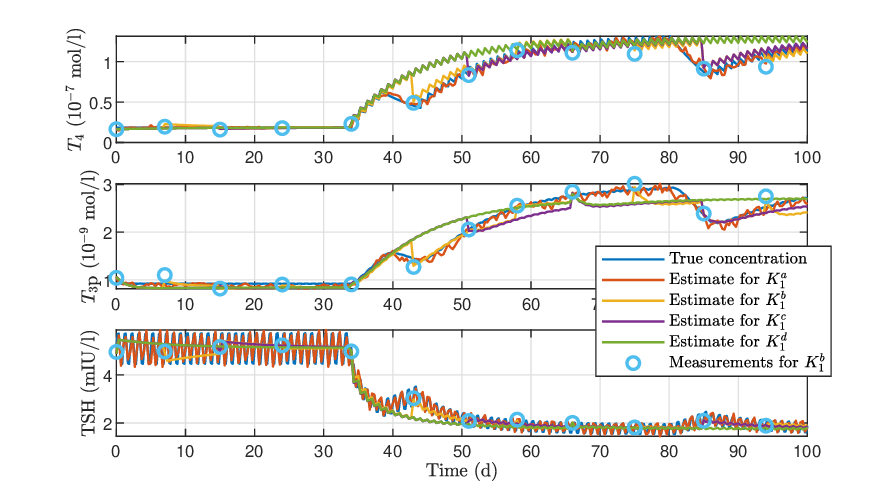}
	\caption{Measured hormone concentrations of virtual patient with hypothyroidism, and estimates, sampling according to $K_1^a$, $K_1^b$, $K_1^c$, and $K_1^d$. }
	\label{fig:hypo-meas}
\end{figure}

\begin{figure}
	\centering
	\includegraphics[width=0.5\textwidth]{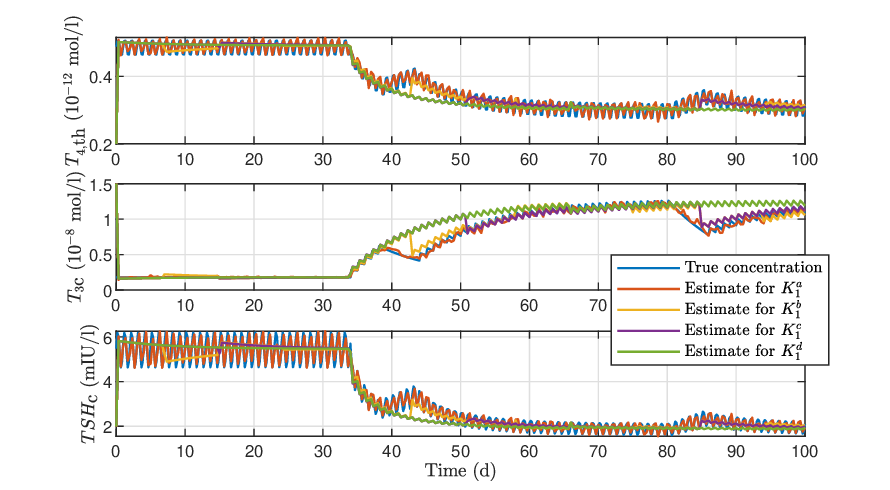}
	\caption{Unmeasured hormone concentrations of virtual patient with hypothyroidism, and estimates, sampling according to $K_1^a$, $K_1^b$, $K_1^c$, and $K_1^d$. }
	\label{fig:hypo-unmeas}
\end{figure}

Plots of the true and estimated states are shown in Figs.~\ref{fig:hypo-meas} and \ref{fig:hypo-unmeas}, the former containing the measured concentrations $T_4$, $T_{3p}$, and $TSH$, and the latter unmeasured concentrations $T_{4,th}$, $T_{3c}$, and $TSH_c$. Until day 34, all states oscillate around consistent values due to $w_{TRH}(t)$. Afterwards, $T_4$, $T_{3p}$, and $T_{3c}$ all increase, and $TSH$, $T_{4,th}$, and $TSH_c$, all decrease, except on days with forgotten dosages, where this is temporarily reverses.
Despite the poor prior $\chi$, estimates obtained via open-loop prediction after the first time step ($8$ \si{\hour}) are near the true states in Fig.~\ref{fig:hypo-unmeas}. This is because $T_4$, $T_{3p}$ and $TSH$ are measured at $t=0$ \si{\hour}, and the dynamics of $T_{4,th}$, $T_{3c}$ and $TSH_c$ have fast transient behavior.
Moreover, the estimates are robustly stable for all sampling schemes, converging to a neighborhood of the true states.
Interestingly, Fig.~\ref{fig:hypo-meas} shows that when sampling based on $K_1^b$, $K_1^c$ and $K_1^d$, the estimates diverge from the true state after measurements.\footnote{This is most clearly seen after day 34, but also occurs beforehand.} This is because the virtual patient is simulated with fixed noise (parameter uncertainty) $w_{G_{D1}}(t)=w_{G_{T3}}(t)=0.1$, but for open-loop prediction between sampled times, we set $w(t)=0$, causing convergence towards a different steady-state trajectory.

\begin{figure}
	\centering
	\includegraphics[width=0.5\textwidth]{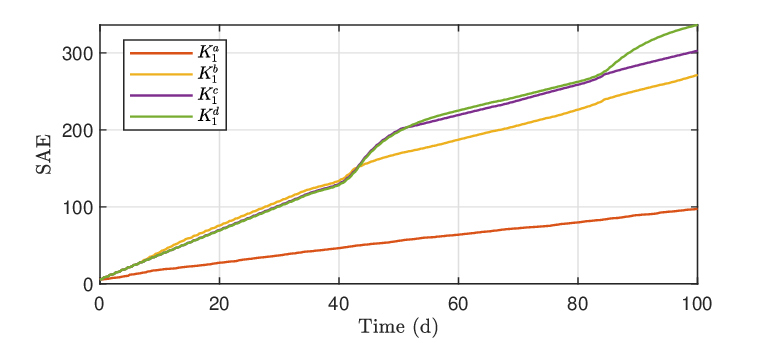}
	\caption{SAE of estimates for virtual patient with hypothyroidism, sampling according to $K_1^a$, $K_1^b$, $K_1^c$, and $K_1^d$. }
	\label{fig:hypo-sae}
\end{figure}

We also plot the sum of absolute errors (SAE) over all states as a function of time---defined as $\mathrm{SAE}(k) = \sum_{i=0}^{k} \Vert \hat{x}_i - x(T_d i) \Vert_1$---in Fig.~\ref{fig:hypo-sae}. 
Before the first forgotten dosages, $K_1^b$ has the worst SAE due to poor measurements on day $7$.  
However, notice that $K_1^b$ samples during both forgotten dosages, and $K_1^c$ samples during the second forgotten dosages, resulting in improved performance at these times, compared to $K_1^d$ which does not sample during either. On the other hand, $K_1^a$ samples at every time step, and therefore not only accounts for the effects of the forgotten dosages, but even picks up intraday oscillations caused by $w_{TRH}(t)$, which can be seen in the $TSH$, $T_{4,th}$, and $TSH_c$ plots across Figs.~\ref{fig:hypo-meas} and \ref{fig:hypo-unmeas}.
These results make sense, since sampling frequently allows more information to be considered by the MHE  algorithm to improve the estimates.
This is supported by the computed root mean square error (RMSE) values (defined as $\mathrm{RMSE} = \sqrt{ \frac{1}{T} \sum_{i=0}^T \Vert \hat{x}_i - x(T_d i) \Vert_2^2 }$ where $T$ is the number of simulated time steps), which were found to be $0.47$ for $K_1^a$, $1.08$ for $K_1^b$, $1.25$ for $K_1^c$, and $1.34$ for $K_1^d$.

\subsection{Hyperthyroidism}
\label{sec:sim-hyper}

Similar to Sec.~\ref{sec:sim-hypo}, for the hyperthyroidism case, the patient initially does not take any medication, then follows a naive oral MMI medication strategy consistent with the guidelines from \cite{ross20162016}. In particular, $7.5$ mg/day is taken if free $T_4$ ($FT_4$) concentration is between $27$--$41$ \si{\pico\mole}/l, $15$ mg/day is taken if $FT_4$ is within $41$--$54$ \si{\pico\mole}/l, and $35$ mg/day is taken if $FT_4$ is above $54$ \si{\pico\mole}/l, with $FT_4$ computed from $T_4$ via \eqref{eqn:ft4} in Appendix~\ref{sec:append-hypo}, and dosages updated when measurements are taken every $4$--$6$ weeks. The corresponding signal $U(\cdot)\in \mathbb{U}_{L\text{-}T}$ is used when solving the MHE problem. The same forgotten dosages and sampling sequences as Sec.~\ref{sec:sim-hypo} are considered.

This scenario was simulated using the model in \eqref{eqn:f-hyper}. Analogously to Sec.~\ref{sec:sim-hypo}, $U(\cdot) \in \mathbb{U}_{MMI}$ is the input, and we set $w_{MMI}(t) = 1 - U_{\text{true}}(t)/U(t)$ if $U(t) > 0$ and $w_{L\text{-}T_4}(t) = 0$ if $U(t)=0$, where $U_{\text{true}}(\cdot) \in \mathbb{U}_{MMI}$ denotes the input if forgotten days are explicitly accounted for.
Moreover, we set $w_{G_{D1}}(t)$, $w_{G_{T3}}(t)$, and $w_{TRH}(t)$, the same as in Sec.~\ref{sec:sim-hypo}.
Measurements are simulated by corrupting $T_4$, $T_{3p}$, and $TSH$, by a noise sampled uniformly over $\mathcal{V}$. We set the initial state as $\begin{bmatrix} 12.45 & 4.68 & 10.57 & 4.28 & 0.86 & 0.92 & 0 \end{bmatrix}^{\top}$, and the MHE prior as $\chi = \begin{bmatrix} 7 & 3 & 7 & 2 & 2 & 2.5 & 1 \end{bmatrix}^{\top}$, with the units ($10^{-12}$ mol/l, $10^{-7}$ mol/l, $10^{-9}$ mol/l, $10^{-8}$ mol/l, mIU/l, mIU/l, $10^{-5}$ mol/l).

\begin{figure}
	\centering
	\includegraphics[width=0.5\textwidth]{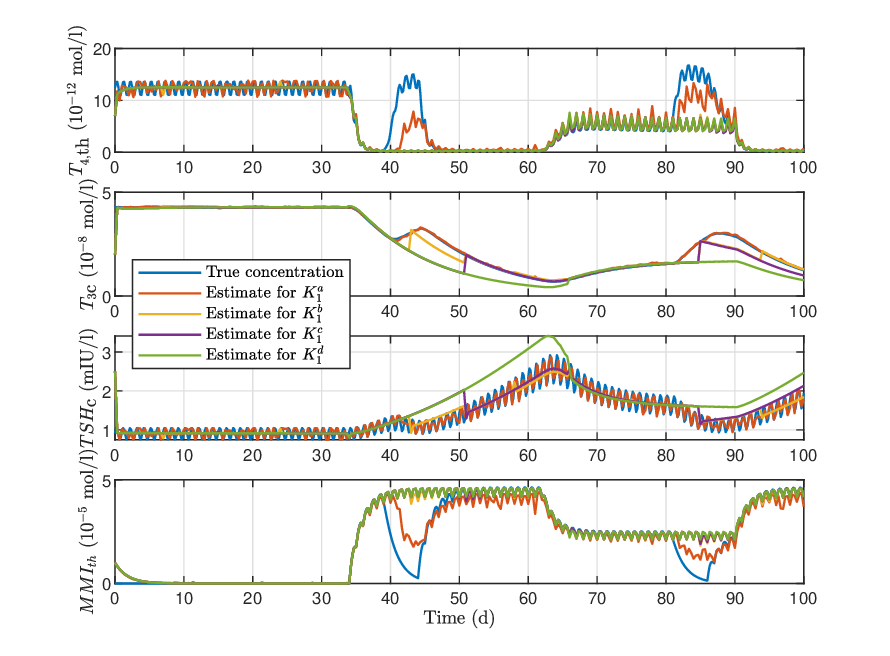}
	\caption{Unmeasured hormone concentrations of simulated patient with hyperthyroidism, and their estimates, when sampling according to $K_1^a$, $K_1^b$, $K_1^c$, and $K_1^d$. }
	\label{fig:graves-unmeas}
\end{figure}

Plots of the true and estimated unmeasured hormone concentrations are shown in Fig.~\ref{fig:graves-unmeas}. Until day 34, all concentrations steadily oscillate around consistent values, then $T_{4,th}$ and $T_{3c}$ decrease, and $TSH_c$ and $MMI_{th}$ both increase, in response to the medication, except when dosages are forgotten. The performance in this scenario is mostly consistent with Sec.~\ref{sec:sim-hypo}, where the estimates are stable with respect to the true state, and improve with more frequent sampling. This can be seen in Fig.~\ref{fig:graves-unmeas}, and is supported by the RMSE values, which were $3.03$ for $K_1^a$, $5.52$ for $K_1^b$, $6.07$ for $K_1^c$, and $6.77$ for $K_1^d$. The most notable difference is that the estimation error is significant during the misreported dosages for the hyperthyroidism case compared to the hypothyroidism case. This is explained by the increased complexity of the hyperthyroidism model, making state estimation more difficult.


\section{Conclusion}

In this work, we evaluated the suitability of PT loop models for state estimation with irregular measurements by verifying sample-based i-IOSS for various sets of sampled time instances, representing potential blood testing schedules in both outpatient and inpatient settings. We implemented sample-based MHE schemes for these models, and tested them on virtual patients treated for hypo- and hyperthyroidism. While estimation performance is robustly stable across all schemes, more frequent sampling decreases the estimation error when misreported dosages occur, allowing tracking of behavior like oscillations operating at faster timescales. 
Clinically, this suggests that a suitable sampling scheme should be chosen based on the certainty in the modeled dynamics and known input.
If there is less certainty (e.g., due to the risk of misreported dosages) then more frequent sampling is recommended to offset the effects of this.
Moreover, intraday sampling is required if accurate estimates throughout a day are important, which could be relevant in patients being treated for more severe forms of hyperthyroidism, such as thyrotoxicosis \citep{ross20162016}.
In future work, we aim to combine the sample-based MHE schemes here with the MPC schemes from \cite{wolff2022optimal} and \cite{wolff2025modeling}. These previous works assume full state measurements, but the results from this paper can be used to make them applicable in realistic settings.
{
\footnotesize
\bibliography{references}             
}               


\appendix

\section{PT loop model for hypothyroidism} \label{sec:append-hypo}

The model $f_{\text{hypo}}$ from \eqref{eqn:f-hypo} is defined such that the following system of equations is satisfied:
{
	\allowdisplaybreaks
	\begin{align}
		&\frac{dT_{4,th}(t)}{dt}= 10^{12} \alpha_{th} \left( G_{T,nom} G_{T,co} \frac{TSH(t)}{TSH(t)+D_T} \right.\\
		& - G_{MCT8} \frac{T_{4,th}(t)}{10^{12} K_{MCT8}+T_{4,th}(t)} \\
		& - G_{D1}(1+w_{G_{D1}}) \frac{T_{4,th}(t) \frac{TSH(t)}{TSH(t)+k_{Dio}}}{T_{4,th}(t) \frac{TSH(t)}{TSH(t)+k_{Dio}} + 10^{12}K_{M1}} \label{eqn:hypo-dynamics-1} \\
		& \left. - G_{D2} \frac{T_{4,th}(t) \frac{TSH(t)}{TSH(t)+k_{Dio}}}{T_{4,th}(t) \frac{TSH(t)}{TSH(t)+k_{Dio}} + 10^{12}K_{M2}} \right) - \beta_{th}T_{4,th}(t) \\
		&\frac{dT_4(t)}{dt} = 10^7 \alpha_T \left ( G_{MCT8} \frac{T_{4,th}(t)}{10^{12}K_{MCT8}+T_{4,th}(t)} \right. \\
		& + u_{L\text{-}T_4}(t) \left(1- w_{L\text{-}T_4}(t)\right) \bigg ) - \beta_T T_4 (t) \label{eqn:hypo-dynamics-2} \\
		&\frac{dT_{3p}(t)}{dt} = 10^9 \alpha_{31} \left ( G_{D1}(1+w_{G_{D1}}) \cdot \right. \\
		& \frac{T_{4,th}(t) \frac{TSH(t)}{TSH(t)+k_{Dio}}}{T_{4,th}(t)\frac{TSH(t)}{TSH(t)+k_{Dio}} + 10^{12}K_{M1}} + G_{D2} \frac{FT_{4}(t)}{FT_4(t)+K_{M2}} +  \\
		& G_{D2} \frac{T_{4,th}(t) \frac{TSH(t)}{TSH(t)+k_{Dio}}}{T_{4,th}(t)\frac{TSH(t)}{TSH(t)+k_{Dio}} + 10^{12}K_{M2}} + G_{D1}(1+w_{G_{D1}}) \cdot  \\
		& \frac{FT_4(t)}{FT_4(t)+K_{M1}} + G_{T3}(1+w_{G_{T3}}) \frac{TSH(t)}{D_t + TSH(t)}  \\
		&  + u_{L-T_3}(t)\left(1- w_{L\text{-}T_3}(t)\right) \Bigg ) - \beta_{31} T_{3p}(t) \label{eqn:hypo-dynamics-3} \\
		&\frac{dT_{3c}(t)}{dt} = 10^8\alpha_{32}G_{D2} \frac{FT_4(t)}{FT_4(t)+K_{M2}} - \beta_{32}T_{3c}(t)  \label{eqn:hypo-dynamics-4}\\
		&\frac{dTSH(t)}{dt} = \frac{\alpha_s G_H TRH(1+w_{TRH})}{TRH(1+w_{TRH}) + D_H} \label{eqn:hypo-dynamics-5} \\
		& \cdot \frac{1}{ ( 1 + S_s \frac{TSH_c(t)}{TSH_c(t) + D_s}  ) (1 + L_s G_R \frac{T_{3N}(t)}{T_{3N}(t)+D_R} ) } - \beta_S TSH(t)  \\
		&\frac{dTSH_c(t)}{dt} = \frac{\alpha_{S2} G_H TRH(1+w_{TRH})}{(TRH(1+w_{TRH}) + D_H)(1 + S_s \frac{TSH_c(t)}{TSH_c(t) + D_s})}  \\
		&  \cdot \frac{1}{ \left ( 1 + L_s G_R \frac{T_{3N}(t)}{T_{3N}(t)+D_R} \right )} - \beta_{S2}TSH_c(t). \label{eqn:hypo-dynamics-6}
\end{align}}

\noeqref{eqn:hypo-dynamics-1}\noeqref{eqn:hypo-dynamics-2}\noeqref{eqn:hypo-dynamics-3}\noeqref{eqn:hypo-dynamics-4}\noeqref{eqn:hypo-dynamics-5}\noeqref{eqn:hypo-dynamics-6}
Here, \begin{align} FT_4 = 10^{-7}T_4 /(1+ K_{41} TBG + K_{42} TBPA), \label{eqn:ft4} \end{align}
$FT_3 = 10^{-9}T_{3p}/({1 + K_{30} TBG})$, and \\ $T_{3N} = 10^{-8}T_{3c} / (1 + K_{31}IBS )$, all with the units mol/l.

Let $\{t_i\}_{i=0}^{\infty}$ satisfying $t_i = 24 \cdot 3600 \cdot i$ represent the time instances where L-T\textsubscript{3} and L-T\textsubscript{4} is taken, in seconds, and let $\{m_{D3,i}\}_{i=0}^{\infty} \in [0,30]^{\infty}$ and $\{m_{D4,i}\}_{i=0}^{\infty} \in [0, 400]^{\infty}$ be the corresponding dosages (in \si{\micro\gram}).
Then, the time-dependent absorption of L-T\textsubscript{3} and L-T\textsubscript{4} are given by
\begin{align}
	&u_{L\text{-}T3}(t)=k_{33}\frac{k_{13}}{k_{13}-(k_{23}+k_{33})} \sum_{ i \in \mathbb{Z}_0 } m_{D3,i}  \mu(t - t_i) \\
	& \quad \cdot \big ( e^{-(k_{23}+k_{33})(t-t_i)} - e^{-k_{13}(t - t_i)} \big ), \label{eqn:lt3}
\end{align}
with the units mol/l/s, and
\begin{align}
	&u_{L\text{-}T4}(t) = k_{34}\frac{k_{14}}{k_{14}-(k_{24}+k_{34})} \sum_{i \in \mathbb{Z}_0} m_{D4,i}\mu(t - t_i) \\
	& \quad \cdot \big( e^{-(k_{24}+k_{34})(t-t_i)} - e^{-k_{14}(t - t_i)} \big ), \label{eqn:lt4}
\end{align}
in mol/l/s, for $t \geq 0$, where $\mu(t):= 0$ for $t < 0$ and $\mu(t) := 1$ for $t \geq 0$ is the Heaviside step function. 
Define
\begin{align}
&\mathbb{U}_{L\text{-}T} := \{ \begin{bmatrix} u_{L\text{-}T3}(\cdot) & u_{L\text{-}T4}(\cdot) \end{bmatrix}^{\top} \mid \eqref{eqn:lt3}, \eqref{eqn:lt4} \text{ satisfied for}\\
& \text{some } \{m_{D3,i}\}_{i=0}^{\infty} \in [0,30]^{\infty} \text{ and } \{m_{D4,i}\}_{i=0}^{\infty} \in [0, 400]^{\infty}\}, \label{eqn:U-hypo}
\end{align}
representing the set of all possible $u_{L\text{-}T_3}$ and $u_{L\text{-}T_4}$ when up to $30$ \si{\micro\gram} and $400$ \si{\micro\gram} of L-T\textsubscript{3} and L-T\textsubscript{4} are taken daily.

\section{PT loop model for hyperthyroidism} \label{sec:append-hyper}
We describe the model of the PT loop for patients with hyperthyroidism undergoing treatment given by
\begin{align}
	\dot{x} = f_{\text{hyper}}(x,u,w), \label{eqn:f-hyper-2}
\end{align}
in more detail compared to Sec.~\ref{sec:model-ct}. The variable $x = \begin{bmatrix} T_{4,th} & T_4 & T_{3p} & T_{3c} & TSH & TSH_c & MMI_{th} \end{bmatrix}^{\top} \in \mathbb{R}_{\geq 0}^7$, with the units ($10^{-12}$ mol/l, $10^{-7}$ mol/l, $10^{-9}$ mol/l, $10^{-8}$ mol/l, mIU/l, mIU/l, $10^{-5}$ mol/l), is the system state, which, compared to Appendix~\ref{sec:append-hypo}, includes $MMI_{th}$, the concentration of MMI in the thyroid. 
The system dynamics $f_{\text{hyper}}$ corresponds to the right-hand side of equations \eqref{eqn:hypo-dynamics-2}--\eqref{eqn:hypo-dynamics-6}, alongside \eqref{eqn:hyper-dynamics-t4th} and \eqref{eqn:hyper-dynamics-mmith}, which are given as follows:
\begin{align}
	&\frac{dT_{4,th}(t)}{dt}=10^{12}\alpha_{th} \left( G_T(t) \frac{TSH(t)}{TSH(t)+D_T} \right.\\
	&\quad - G_{MCT8} \frac{T_{4,th}(t)}{10^{12}K_{MCT8}+T_{4,th}(t)} \\
	&\quad - G_{D1}(1+w_{G_{D1}}) \frac{T_{4,th}(t) \frac{TSH(t)}{TSH(t)+k_{Dio}}}{T_{4,th}(t) \frac{TSH(t)}{TSH(t)+k_{Dio}} + 10^{12}K_{M1}} \\
	& \quad \left. - G_{D2} \frac{T_{4,th}(t) \frac{TSH(t)}{TSH(t)+k_{Dio}}}{T_{4,th}(t) \frac{TSH(t)}{TSH(t)+k_{Dio}} + 10^{12}K_{M2}} \right) \\
	& \quad - \beta_{th}T_{4,th}(t),  \label{eqn:hyper-dynamics-t4th}
\end{align}
where $G_T(t) = G_{T,nom} G_{T,co} TPO_a(t)$, $TPO_a(t) = c_0 (1 + \exp{(-c_1 ( -(10^{-5}MMI_{th}(t))^{1/c_2} +c_3 ))})^{-1}$, and
\begin{align}
	&\frac{dMMI_{th}(t)}{dt} = 10^5u_{MMI}(t)(1 - w_{MMI}(t)) \\
	& \quad - \beta_{M,th}MMI_{th}(t). \label{eqn:hyper-dynamics-mmith}
\end{align}
The parameters are from Table~\ref{tab:params-fixed}, except that we choose $G_{T,co}=7$, representing an overactive thyroid.
This model is obtained by modifying the one from \cite{wolff2025modeling} to consider process noise $w = \begin{bmatrix} w_{G_{D1}} & w_{G_{T3}} & w_{TRH} & w_{MMI}\end{bmatrix}^{\top} \in \mathbb{R}^4$. 
We refer to \cite{wolff2025modeling} for further details about the model, and focus on describing our modifications.
The variables $w_{G_{D1}}$, $w_{G_{T3}}$, and $w_{TRH}$, are the same as the hypothyroidism case. However, $w_{MMI}$ captures uncertainty in knowledge of $u_{MMI}$ potentially caused by misreported medication, such that compared to \cite{wolff2025modeling}, \eqref{eqn:hyper-dynamics-mmith} is obtained by multiplying $u_{MMI}$ with $(1 - w_{MMI})$.
The known, time-varying input is $u(t) = u_{MMI}(t) \in \mathbb{R}_{\geq 0}$, which describes the time-dependent absorption of MMI from the plasma to the thyroid (in mol/l/s). This can be computed given historical knowledge of the MMI dosages taken by the patient as follows:
\begin{align}
	u_{MMI}(t) = \alpha_{M,th} G_{M,th} \frac{MMI_{Pl}(t)}{K_{M,th}+MMI_{Pl}(t)}. \label{eqn:u-mmipl}
\end{align}
Here, $MMI_{Pl}$ denotes the concentration of MMI in the plasma. To define it, let $\{t_i\}_{i=0}^{\infty}$ satisfying $t_i = 24 \cdot 3600 \cdot i$ represent the time instances where MMI is orally taken, in seconds, and let $\{m_{MMI,i}\}_{i=0}^{\infty} \in [0,35]^{\infty}$ be the corresponding dosages (in \si{\milli\gram}).
Then, $MMI_{Pl}$ is given by
\begin{align}
	& MMI_{Pl}(t)= \frac{f_b k_a}{V(k_a - k_e)} \sum_{i \in \mathbb{Z}_0} m_{MMI,i} \mu(t - t_i) \\
	& \quad \cdot  (e^{-k_e(t - t_i)} - e^{-k_a(t - t_i)}). \label{eqn:mmipl-oral}
\end{align}
For convenience, we also define
\begin{align}
	\mathbb{U}_{MMI} := \{ u_{MMI}(\cdot) \mid \text{\eqref{eqn:u-mmipl} is satisfied} \\
	\text{for some } \{m_{MMI,i}\}_{i=0}^{\infty} \in [0,35]^{\infty} \}, \label{eqn:U-mmi}
\end{align}
representing the set of all possible $u_{MMI}(\cdot)$ when up to $35$ \si{\milli\gram} of MMI is taken daily.

\newpage

\section{Parameters}
\label{sec:params}

All model parameters are in Table~\ref{tab:params-fixed}.

\begin{table}[htbp]
	\centering
	\caption{Fixed parameters across all models.}
	\label{tab:params-fixed}
	\renewcommand{\arraystretch}{1.2}
	\resizebox{0.48\textwidth}{!}{
	\begin{tabular}{cc|cc}
		\hline
		{Parameter} & {Value} & {Parameter} & {Value} \\
		\hline
		$TBG$ & $300$ nmol/l & $TBPA$ & $4.5$ \si{\micro\mole}/l \\
		$IBS$ & $8$ \si{\micro\mole}/l & TRH & $6.9$ nmol/s \\
		$G_H$ & $817$ mIU/s & $D_H$ & $47$ nmol/s \\
		$\alpha_S$ & $0.4$ l$^{-1}$ & $\beta_S$ & $2.3\cdot10^{-4}$ s$^{-1}$ \\
		$L_S$ & $1.68$ l/\si{\micro\mole} & $G_{T,nom}$ & $3.4$ pmol/s \\
		$D_T$ & $2.75$ mIU/l & $\alpha_T$ & $0.1$ l$^{-1}$ \\
		$\beta_T$ & $1.1\cdot10^{-6}$ s$^{-1}$ & $K_{M1}$ & $500$ nmol/l \\
		$\alpha_{31}$ & $2.6\cdot10^{-2}$ l$^{-1}$ & $\beta_{31}$ & $8\cdot10^{-6}$ s$^{-1}$ \\
		$G_{D2}$ & $4.3$ fmol/s & $K_{M2}$ & $1$ nmol/l \\
		$\alpha_{32}$ & $1.3\cdot10^{-5}$ l$^{-1}$ & $\beta_{32}$ & $8.3\cdot10^{-4}$ s$^{-1}$ \\
		$\alpha_{S2}$ & $2.6\cdot10^{5}$ l$^{-1}$ & $\beta_{S2}$ & $140$ s$^{-1}$ \\
		$D_R$ & $100$ pmol/l & $G_R$ & $1$ mol/s \\
		$S_S$ & $100$ l/mIU & $D_S$ & $50$ mIU/l \\
		$K_{30}$ & $2\cdot10^{9}$ l/mol & $K_{31}$ & $2\cdot10^{9}$ l/mol \\
		$K_{41}$ & $2\cdot10^{10}$ l/mol & $K_{42}$ & $2\cdot10^{8}$ l/mol \\
		$\alpha_{th}$ & $250$ l$^{-1}$ & $\beta_{th}$ & $4.4\cdot10^{-6}$ s$^{-1}$ \\
		$k_{Dio}$ & $1$ mIU/l & $K_{MCT8}$ & $4.7\cdot10^{-6}$ mol/l \\
		$G_{D1}$ & $1.98\cdot10^{-8}$ mol/s & $G_{T3}$ & $23.1\cdot10^{-13}$ mol/s \\
		$G_{MCT8}$ & $1.94\cdot10^{-6}$ mol/s & $c_0$ & $0.97$ \\
		$G_{M,th}$ & $1.92\cdot10^{-12}$ mol/s & $c_1$ & $1.47\cdot10^{4}$ \\
		$K_{M,th}$ & $7.28\cdot10^{-7}$ mol/l & $c_2$ & $1.36\cdot10^{4}$ \\
		$\alpha_{M,th}$ & $250$ l$^{-1}$ & $c_3$ & $3.23\cdot10^{4}$ \\
		$\beta_{M,th}$ & $6.42\cdot10^{-6}$ s$^{-1}$ & $f_b$ & $0.93$ \\
		$k_a$ & $1.02$ l/h & $k_e$ & $0.106$ l/h \\
		$V$ & $281$ &  &  \\
		\hline
	\end{tabular}}
\end{table}







\end{document}